\documentclass[11pt]{article}
\usepackage[utf8]{inputenc}
\usepackage[T1]{fontenc}
\usepackage[a4paper,top=1in,bottom=1in,left=1in,right=1in,marginparwidth=1.75cm]{geometry}
\usepackage{indentfirst}
\usepackage{graphicx}
\usepackage{setspace}
\usepackage[hyphens]{url}
\usepackage{listings}
\usepackage{pgfplots}
\usepgfplotslibrary{fillbetween}
\pgfplotsset{compat=1.16,width=10cm}
\usepackage[capposition=top]{floatrow}
\usepackage{adjustbox}
\usepackage{cool}
\usepackage{comment}

\usepackage[hyphens]{url}
\usepackage{amsmath}
\newcommand{\citeapos}[1]{\citeauthor{#1}{\textcolor{bluepigment}{'s }}\citeyearpar{#1}}\usepackage{pifont}
\usepackage{pifont}
\usepackage{pifont}
\newlength{\bibitemsep}
\newlength{\bibparskip}
\let\oldthebibliography\thebibliography
\renewcommand\thebibliography[1]{
 \oldthebibliography{#1}
  \setlength{\parskip}{\bibitemsep}
  \setlength{\itemsep}{\bibparskip}
}
\usepackage[super]{nth}

\usepackage[capposition=top]{floatrow}
\usepackage{physics}
\usepackage[font=large,labelfont=bf]{caption}
\usepackage{csquotes}
\usepackage{multirow}

\usepackage[normalem]{ulem}
\useunder{\uline}{\ul}{}
\usepackage{wrapfig, framed, caption}
\usepackage{mathrsfs}
\usepackage{flafter}
\definecolor{bluepigment}{rgb}{0.2, 0.2, 0.6}
\usepackage{listings}
\usepackage[bottom, flushmargin]{footmisc}
\usepackage{subfig}

\usepackage{enumitem}

\makeatletter
\def\footnoterule{\kern-8\p@
  \hrule \@width 2in \kern 7.6\p@}
\newcommand*\bigcdot{\mathpalette\bigcdot@{.5}}
\newcommand*\bigcdot@[2]{\mathbin{\vcenter{\hbox{\scalebox{#2}{$\m@th#1\bullet$}}}}}
\makeatother

\usepackage{hyperref}
\hypersetup{colorlinks,allcolors=bluepigment}
\usepackage{natbib}

\title{\textbf{Out of the Loop: \\ How Dangerous is Weaponizing \\ Automated Nuclear Systems?}\thanks{  We thank Sanghyun Han, Jenna Jordan, David Logan, Scott Sagan, participants of the MIT Security Studies Working Group, the Georgia Tech STAIR Workshop, the Carnegie Mellon Political Science Research Workshop, and the 2024 International Studies Association Annual Conference for helpful comments and advice. We are also deeply grateful to Graham Elder, Isabel Leong, and Stotra Pandya for valuable research assistance.}}
\author{Joshua A. Schwartz\thanks{Assistant Professor of International Relations and Emerging Technology, Carnegie Mellon Institute for Strategy and Technology, Carnegie Mellon University, \href{mailto:joshschwartz@cmu.edu}{joshschwartz@cmu.edu.}}\hspace{2mm} \& Michael C. Horowitz\thanks{Richard Perry Professor of Political Science and Director of Perry World House, Department of Political Science, University of Pennsylvania, \href{mailto:horom@sas.upenn.edu}{horom@sas.upenn.edu.}}
}
\date{\today}

\begin{document}

\maketitle

\thispagestyle{empty} 

\vspace{-8mm}

{\centering{\section*{Abstract}}}

\noindent Are nuclear weapons useful for coercion, and, if so, what factors increase the credibility and effectiveness of nuclear threats? While prominent scholars like Thomas Schelling argue that nuclear brinkmanship, or the manipulation of nuclear risk, can effectively coerce adversaries, others contend nuclear weapons are not effective tools of coercion, especially coercion designed to achieve offensive and revisionist objectives. Simultaneously, there is broad debate about the incorporation of artificial intelligence (AI) into military systems, especially nuclear command and control. We develop a theoretical argument that explicit nuclear threats implemented with automated nuclear launch systems are potentially more credible compared to ambiguous nuclear threats or explicit nuclear threats implemented via non-automated means. By reducing human control over nuclear use, leaders can more effectively tie their hands and thus signal resolve. While automated nuclear weapons launch systems may seem like something out of science fiction, the Soviet Union deployed such a system during the Cold War and the technology necessary to automate the use of force has developed considerably in recent years due to advances in AI. Preregistered survey experiments on an elite sample of United Kingdom Members of Parliament and two public samples of UK citizens provide support for these expectations, showing that, in a limited set of circumstances, nuclear threats backed by AI integration have credibility advantages, no matter how dangerous they may be. Our findings contribute to the literatures on coercive bargaining, weapons of mass destruction, and emerging technology.

\clearpage
\pagenumbering{arabic}
\section*{Introduction}
\doublespacing

The ``Doomsday Machine:'' A nuclear arsenal programmed to automatically explode should the Soviet Union come under nuclear attack. ``Skynet:'' An artificial general superintelligence system that becomes self-aware and launches a series of nuclear attacks against humans to prevent it from being shut down. ``War Operation Plan Response (WOPR):'' A supercomputer that is given access to the US nuclear arsenal, programmed to run continuous war games, and comes to believe a simulation of global thermonuclear war is real. These systems appear in the movies \textit{Dr. Strangelove}, \textit{The Terminator}, and \textit{WarGames}. However, the idea of automating nuclear weapons use---that is, creating a computer system that, once turned on, could launch nuclear weapons on its own without further human intervention \citep[][21]{us2023dod}---is not just a Hollywood plot device. For example, the Soviet Union developed a semi-autonomous system called ``Dead Hand'' or ``Perimeter'' during the Cold War \citep{thompson2009inside}. Once the system was turned on, it monitored for evidence of a nuclear attack against the Soviet Union using a network of radiation, seismic, and air pressure sensors. Upon detecting that a nuclear weapon had been exploded on Soviet territory, it would check to see if there was an active communications link to the Soviet leadership. If the communications link was inactive, Dead Hand would assume Soviet leadership was dead and transfer nuclear launch authority to a lower-level official operating the system in a protected bunker. Today, the technology related to automating the use of lethal force has advanced significantly beyond what was possible in the Cold War due to advances in artificial intelligence \citep{scharre2018army, horowitz2019speed}. 

How the possibility of greater automation in warfare, especially in the realm of nuclear command and control \citep[e.g.,][]{feaver1992command}, will impact international politics is particularly relevant today. In recent years, norms against using weapons of mass destruction have been challenged \citep{tannenwald2018strong, price2019syria}. The Syrian, North Korean, and Russian governments have used chemical weapons on the battlefield or to carry out assassinations. Vladimir Putin and other Russian officials, such as former president and prime minister Dmitry Medvedev, have threatened nuclear use in the context of the Russia-Ukraine War \citep{mills2023russia}. Donald Trump implicitly threatened nuclear use against North Korea in public with his infamous ``fire and fury'' and ``totally destroy North Korea'' comments. He also explicitly discussed nuclear use in private with his advisors \citep{schmidt2023donald}. China is making large-scale investments in expanding their nuclear arsenal by several magnitudes, and countries around the world are reconsidering their commitment to nuclear nonproliferation in light of changes to US foreign policy being implemented by the second Trump administration. Some scholars have even advocated for the US to develop its own Dead Hand system \citep{lowther2019america}.

We assess the impact of automation in the nuclear realm by focusing on nuclear coercion. That is, the \textit{threatened} use of nuclear weapons to deter actors from changing the status quo or compelling them to alter it \citep{pauly2022psychology}. We specifically study nuclear brinksmanship, which involves a ``competition in risk taking, characterized not so much by tests of force as by tests of nerve'' \citep[][94]{schelling1966arms}. Before the nuclear revolution, warfare may have been more aptly characterized as a competition in military capabilities. Nuclear weapons changed the character of warfare because of their destructive power \citep{powell1990nuclear}. Actually \textit{using} nuclear weapons against another nuclear-armed state with a secure second-strike capability may be irrational since it comes with a very high risk of retaliation and complete destruction. On the other hand, \textit{threatening} to use nuclear weapons for the purpose of coercing your adversary and taking active steps to increase the risks of nuclear use could be rational because it is a strategy that enables a state to potentially achieve their objectives without having to actually resort to nuclear war \citep{kroenig2018logic}. 

This research agenda intersects with two foundational debates in political science. First, are nuclear weapons useful for coercion at all, and, if so, then what factors make the threatened use of nuclear weapons more credible? Although nuclear weapons do appear to be effective at deterring major nuclear attacks against a country’s homeland due to the likelihood of retaliation and mutually assured destruction, scholars vigorously disagree about their coercive utility beyond these fundamentally defensive scenarios. While some argue that nuclear weapons can be effective tools of coercion \citep{kroenig2018logic}, others contend that ``despite their extraordinary power, nuclear weapons are uniquely poor instruments of compellence'' \citep[][173]{sechser2017nuclear}. Relatedly, scholars debate when nuclear threats are more or less credible. Many potential answers to this question have been proposed, from nuclear superiority relative to an opponent \citep{kroenig2018logic, fanlo2023disadvantage} and putting nuclear forces on hair-trigger alert \citep{sagan1985nuclear} to leader madness \citep{ellsberg1959political, mcmanus2019revisiting, schwartz2023madman}. However, much of the available evidence on the efficacy of nuclear coercion comes from case studies and observational statistical analyses, which can suffer from internal validity challenges, or from experiments conducted on the general public, who have an indirect impact on a country's foreign policy. Whether automation in the nuclear realm might have a similar impact on coercive efficacy as the above strategies has also been understudied. 

The second debate our project overlaps with is whether emerging technologies---such as autonomous systems and artificial intelligence (AI)---aid or detract from strategic stability \citep{sechser2019emerging, onderco2021emerging, schneider2023hacking}. Some scholars contend that these technologies are destabilizing because they, for example, make decisions incredibly quickly and thus escalation can occur before humans can step in to control it \citep{chyba2020new, johnson2020artificial, futter2022disruptive}. By contrast, others argue autonomous systems and AI can have stabilizing effects \citep{horowitz2019stable, cox2021unavoidable}. For instance, effective systems may provide more accurate and timely early warning and detection of nuclear attacks, buying decision-makers time to carefully consider a response. The effects of automation on the use of nuclear weapons for coercive purposes, however, remains mostly unexplored. 


We contribute to these debates by theorizing that, based on logic that goes back to \citet{schelling1966arms}, greater automation in the nuclear realm---while incredibly dangerous and not without significant downsides---can enhance the credibility of nuclear threats and better enable states to win games of nuclear brinksmanship in some cases. The logic of our argument is simple. Unless it is integrated into their programming, computers do not care that actually using nuclear weapons in a particular situation may be irrational or enormously costly. Consequently, by developing and activating an automated nuclear weapons system, leaders can more credibly threaten to use nuclear weapons compared to when it is a human that retains complete control over the process \citep{pauly2022psychology}. Like other strategies, such as public threats that put a leader's reputation on the line and launch-on-warning policies, automated nuclear weapons systems enable leaders to more effectively ``tie their hands,'' increase nuclear risks, and throw the proverbial steering wheel out of the car in a game of chicken \citep{fearon1997signaling, yarhi2018, horowitz2019speed}. 

We test our expectations with preregistered experiments on an elite sample of UK Members of Parliament (MPs) and two representative samples of the UK public. The experimental scenario involves a future Russian invasion of Estonia. We then randomly assign respondents to receive different types of Russian threats warning NATO countries, including the UK, to avoid taking certain kinds of military actions to aid Estonia. Our experiments yield two primary findings. First, explicit nuclear threats---implemented either via automated or non-automated means---are both more credible (in terms of the perceived likelihood that nuclear weapons will be used if the red line is violated) and more effective (in terms of respondents' willingness to give in to the threat) than ambiguous threats that do not explicitly threaten nuclear use. Second, all three studies yield at least some evidence---albeit somewhat inconsistent evidence---that nuclear threats implemented via automation are more credible and/or effective than nuclear threats implemented via non-automated means. Troublingly, these dynamics suggest that nuclear-armed states may have incentives to automate elements of nuclear commanad and control to make their threats more credible. 


The results also show that automated nuclear threats can---in some ways---backfire for the foreign policy of the coercing country. Among our public sample, we find that nuclear threats implemented via automation increase threat perceptions of a country and make the targeted population more supportive of increasing military spending and less supportive of nuclear disarmament compared to ambiguous threats that do not explicitly threaten nuclear use. We also find that members of the public---but, interestingly, not policymakers---are significantly more likely to believe that threats implemented via automation will lead to accidental nuclear use compared to threats implemented without automation. On the one hand, this belief could discourage potential coercers from adopting automated systems because of fear of losing control. On the other hand, since the goal of nuclear brinksmanship is to increase the perception of risk in the mind of the target, this dynamic may make nuclear coercion more effective. In any case, our findings indicate that just as other military innovations---such as gunpowder, railroads, aircraft, and precision-guided munitions---changed the character of warfare, so might the integration of automated and AI technologies with nuclear systems. 

This study makes several key contributions. First, the results move the debate forward about whether and under what conditions nuclear weapons have coercive utility. Our project indicates that nuclear weapons can indeed provide states leverage in international disputes. Automation is a technological tool and explicitness is a rhetorical tool that states can potentially exploit to increase the credibility and effectiveness of nuclear threats. Second, while automated nuclear systems may have the potential to be stabilizing, our project suggests they can also be effectively used in highly destabilizing ways, such as offensive and revisionist coercive efforts. Third, given the lack of real-world historical data on the use of automated systems in nuclear coercion, our experimental approach provides unique insights since we can make more credible causal inferences. Experiments in general offer an effective means with which to build micro-foundations, and they have been under-utilized to address the two principal debates we highlight. Fourth, this paper speaks to broader debates about how advances in AI and autonomous systems will shape warfare, given that most military applications will occur outside the nuclear realm. If automated threats generate credibility in the nuclear realm, it may suggest something about conventional deterrence and coercion as well, which is a potential avenue for future research. Although we find that automating nuclear use can potentially be useful for coercion, this does \textit{not} at all suggest that states should adopt these systems. Doing so has the potential to increase the risk of nuclear use and make dangerous security dilemmas more likely. However, states should prepare for the possibility that their adversaries might consider automating elements of nuclear command and control if they believe it will increase their coercive leverage.


\section*{Debates in the Literature}

\subsection*{The Debate Over Nuclear Weapons and Coercion}

Sun Tzu famously said, ``For to win one hundred victories in one hundred battles is not the acme of skill. To subdue the enemy without fighting is the acme of skill.'' The logic of this argument is that actually fighting costs blood and treasure, whereas achieving your goals without using military force preserves lives, money, and other resources like land and weapons. One way to achieve foreign policy objectives without fighting is to utilize coercion, which involves the \textit{threat} to do something---such as use military force---in the future. Coercion can either be utilized to (a) deter, which involves preventing a target from taking an action, or to (b) compel, which entails convincing a target to take an action. The latter type of coercion is generally accepted as more difficult to achieve because, in that case, it will be obvious a target ``gave in'' to the threat, which causes the kind of embarrassment leaders typically wish to avoid. Threats can also involve either punishment (e.g., ``if you attack my country, then I will hurt your country'') or denial (e.g., ``you can try to attack my country, but I will prevent you from successfully doing so''). 

Although there are many factors that impact the efficacy of coercive efforts, a critical one is a state's military capabilities. For example, the greater a state's capacity to engage in deadly violence, the more pain they can realistically threaten. Nuclear weapons, then, would ostensibly confer significant coercive leverage given their immense destructive capabilities. As \citet[][9]{pape1996bombing} said, ``Nuclear weapons can almost always inflict more pain than any victim can withstand.'' However, for coercion to achieve its desired goal (i.e., for it to be \textit{effective}) the target must believe that if they violate your threat, then they will face the promised punishment (i.e., your threat must be \textit{credible}). Or, at least, the target must believe there is an unacceptably high risk of the punishment being imposed. The question of whether nuclear threats are credible has led to two fundamental debates among scholars that are germane to this project. First, are nuclear weapons useful at all for coercion, especially coercion that does not involve deterrence and self-defense? Second, if nuclear weapons can aid in coercion, then what factors make the threatened use of nuclear weapons more credible? 

With respect to the first question, there are two schools of thought that can broadly be labeled as nuclear coercion skeptics and nuclear coercion believers. Skeptics question whether the threat of nuclear use by State A against State B is credible outside a scenario where State B has launched a significant, direct attack against State A's territory \citep[e.g.,][]{sechser2017nuclear}. In other words, besides for direct self-defense against a potentially existential attack, the threat of nuclear use may lack credibility for several reasons. 

First, when facing a nuclear-armed opponent with a secure second-strike capability, using nuclear weapons may simply be irrational because it has a high chance of leading to devastating nuclear retaliation \citep{jervis1989meaning}. In accordance with this logic, some studies find that nuclear-armed states are less likely to go to war \citep{rauchhaus2009evaluating} and crises that involve nuclear-armed states are more likely to end without violence \citep{asal2007proliferation}.\footnote{\citet{bell2015questioning} question results like these using different, and arguably more appropriate, methodologies. Other scholars also assess the conditions under which nuclear weapons deter \citep{narang2014nuclear}.} The fact that the Cold War between the United States and Soviet Union stayed cold and never escalated into World War III is also frequently cited as illustrating the pacifying effects of nuclear weapons. 

Second, even against non-nuclear states that cannot threaten nuclear retaliation, nuclear threats may also lack credibility because the political, economic, and social costs of carrying out the threat would be extremely high \citep{sechser2017nuclear}. Some scholars argue there is a global nuclear taboo, which is ``a de facto prohibition against the use of nuclear weapons...[that] involves expectations of awful consequences or sanctions to follow in the wake of a taboo violation'' \citep[][436]{tannenwald1999nuclear}.\footnote{Other scholars argue there exists a ``tradition'' of nuclear non-use, which is weaker than a ``taboo'' \citep{paul2020tradition}.} For example, a state that uses nuclear weapons might face military retaliation, devastating economic sanctions, a loss of reputation and international standing, and long-term efforts to balance against it. Leaders may also decline to use nuclear weapons against non-nuclear states due to their own moral compunctions. This logic may help explain why countries have not used nuclear weapons even when they possess an asymmetric nuclear edge (e.g., the United States' war in Vietnam). More systematically, using statistical analysis and case studies, \citet{sechser2017nuclear} find little evidence that nuclear weapons are useful for more than self-defense. The reason, they conclude, ``is that it is exceedingly difficult to make coercive nuclear threats believable. Nuclear blackmail does not work because threats to launch nuclear attacks for offensive political purposes fundamentally lack credibility'' \citep[][236]{sechser2017nuclear}.

Third, making nuclear threats credible might be particularly challenging in cases of extended nuclear deterrence, which involve implicit or explicit threats to respond to an attack against an ally with nuclear weapons \citep{huth1988extended}. For example, French President Charles de Gaulle famously questioned whether the United States would really ``trade New York for Paris'' \citep{FRUS}. Making nuclear threats credible in this context may be more difficult than making conventional threats credible, as carrying out the former would likely result in more severe retaliation against a country's home territory than carrying out the latter. 

On the other hand, nuclear coercion believers contend that nuclear threats can be made credible, even outside the context of deterrence and self-defense. One key argument that purportedly helps solve the problem of credibility is \citeapos{schelling1980strategy} theory of nuclear brinksmanship, or the ``threat that leaves something to chance.'' Although actually using nuclear weapons against a nuclear-armed state with a second-strike capability is illogical due to the dynamics of mutually assured destruction (MAD), states may be able to take certain actions to increase the \textit{risk} of nuclear use and thus convince adversaries that nuclear threats could end up being acted upon. A classic example Schelling utilizes to illustrate brinksmanship is a situation in which two mountain climbers are tied together. If one climber wanted to coerce the other into doing something, they could threaten to jump off the mountain. While this might not be a credible threat from one perspective since jumping off the mountain would cause both climbers to die, if the one climber moves closer to the edge of the cliff or takes one foot off the mountain, then the risks of a devastating accident would increase substantially and the other climber might decide to give in to avoid disaster. Therefore, even compellent, offensive, and revisionist nuclear threats could potentially be made more credible by using similar means.

This paper directly contributes to this first debate by theorizing and empirically testing whether nuclear weapons can be useful for coercion in a case that involves clear offensive objectives rather than self-defense. It contributes to the second debate as well about the factors that make the threatened use of nuclear weapons more credible. Several possibilities have been raised. First, states can delegate the authority to use nuclear weapons to lower-level military commanders. Doing so raises the risk that ``rogue military officers could take matters into their own hands and release nuclear weapons'' \citep[][14]{narang2014nuclear}. In other words, giving the authority to launch nuclear weapons over to an entity that is less likely to be dissuaded by the risks of MAD can make nuclear threats more credible \citep{pauly2022psychology}. For example, the Soviet Dead Hand system involved delegating authority to lower-level military commanders in the event of a crisis, and US presidents from Eisenhower to Reagan delegated nuclear launch authority in various circumstances. This policy option is closely related to this project, which examines the impact of delegating nuclear launch authority to a machine rather than another human.

Second, leaders can make public threats that are designed to engage audience costs and put their reputation on the line. Making public threats can help tie leaders' hands and signal resolve because if they fail to follow through on the threat, then they will pay inconsistency costs among their domestic public for saying one thing and doing another \citep{kertzer2016decomposing}. In support of this general argument, \citet{fuhrmann2014signaling} find that when nuclear states sign defense pacts with other countries---a kind of public commitment---it reduces the chance of those countries being targeted in violent militarized disputes. 

An unanswered question related to audience costs and threat credibility is how explicit nuclear threats must be to confer coercive leverage. One group of scholars argue that ambiguous nuclear threats, or even the mere presence of nuclear weapons, may enhance coercion. For example, \citet{sagan2000commitment} contends that vague nuclear threats to deter chemical and biological weapons use creates a ``commitment trap'' that may enhance credibility, but also means leaders will face domestic political punishment if they do not use nuclear weapons when their red lines are violated. Thus, leaders may become locked-in to their positions and face strong political incentives to launch nuclear attacks even when doing so is not the optimal policy decision. \citet[][351]{kissinger1955force} similarly said, ``overt threats have become unnecessary; every calculation of risks will have to include the Soviet stockpile of atomic weapons and ballistic missiles.'' The mainstream Chinese view is also that Russia's largely ambiguous nuclear signaling has worked to deter Western escalation without the need for explicit nuclear threats \citep{zhao2022implications}.

However, another group of scholars argue that explicit nuclear threats engage audience costs more than ambiguous threats and thus are more likely to lead to coercive leverage \citep{snyder2011cost}. \citet{smetana2023commitment} conduct survey experiments on the US public and find that leaders are not punished for failing to uphold ambiguous nuclear threats, but are punished when they do not follow through on explicit nuclear threats. In other words, clear, unambiguous nuclear threats generate greater audience costs and there is no ``commitment trap'' associated with vague nuclear threats. While leaders have historically been hesitant to make these kinds of explicit threats for fear that it will reduce their future flexibility, from a theoretical perspective ``maximum explicitness and clarity in threats tends to produce maximum credibility'' \citep[][216]{snyder2015conflict}. Nevertheless, prior work has not empirically tested the coercive efficacy of implicit or ambiguous nuclear threats relative to more explicit threats. We provide a first step towards addressing that gap in the literature.\footnote{Although the study by \citet{smetana2023commitment} provides suggestive evidence that explicit nuclear threats are likely to be more effective in coercive bargaining than ambiguous ones, it focuses on the domestic reaction to these threats rather than the reaction of the target state. The latter is most important when assessing threat credibility.}  




There are several other factors that might make the use of nuclear weapons more credible. For instance, nuclear superiority relative to an opponent (i.e., having a greater quantity and/or quality of nuclear weapons) may enhance bargaining leverage by increasing a state's willingness to engage in risky behavior \citep{kroenig2018logic}. States with nuclear superiority know if a nuclear war does occur, then they at least will suffer relatively less damage than their opponent.\footnote{Other work, however, questions the empirical validity of this argument \citep{logan2022nuclear, suh2023nuclear}.} Leaders can also increase the risk of accidents by putting their nuclear forces on hair-trigger alert \citep{sagan1985nuclear}, or adopting the so-called ``madman strategy'' to try and convince their adversaries that even if acting on nuclear threats is irrational, they just might be crazy enough to do so \citep{ellsberg1959political, mcmanus2019revisiting, schwartz2023madman}. 




While all of these strategies can increase risks by leaving ``something to chance,'' at the end of the day, a human being still has to press the nuclear button. As \citet[][12]{pauly2022psychology} said, ``Barring a preexisting doomsday machine, leaders still must make a conscious choice to use nuclear weapons, even in response to an attack that is assumed to be so provocative as to demand one.'' Nevertheless, it is precisely the potential use of a doomsday machine, as in \textit{Dr. Strangelove} or like the Soviet's Dead Hand system, that we are interested in and turn to now.

\subsection*{The Debate Over Automation and Strategic Stability}

Given the unthinkable destructiveness a nuclear war would bring, a key concern of scholars and policymakers is maintaining strategic stability. Strategic stability can be defined narrowly as the lack of incentives for any country to attempt a disarming nuclear first strike or, more broadly, as the lack of military conflict between nuclear-armed states \citep[][101]{boulanin2020artificial}. Given the use of automated systems (such as Dead Hand) in the past, as well as emerging technologies (such as AI), one key debate is whether these technologies will enhance or undermine strategic stability. We can derive two broad schools of thought from the writing above and the existing literature---nuclear automation optimists and pessimists. Though very few scholars advocate automating nuclear use, there is a potential nuclear automation optimist argument for why it could bolster strategic stability. Of course, some scholars articulate arguments that are consistent with both schools or have more nuanced views.

The automation optimism argument relies on several points. First, new technologies---such as hypersonic weapons, stealthy nuclear-armed cruise missiles, and underwater nuclear-armed drones---substantially compress the time available to respond to a nuclear first strike, making such an attack more likely to succeed and undermining strategic stability \citep{lowther2019america}. To avoid a scenario where a first strike cripples a country's nuclear command, control, and communications systems (NC3) before they are able to retaliate, countries can adopt automated nuclear systems (like Dead Hand) to help ensure retaliation and enhance nuclear deterrence. As \citet{lowther2019america} argue, ``To maintain the deterrent value of America’s strategic forces, the United States may need to develop something that might seem unfathomable---an automated strategic response system based on artificial intelligence.''

Second, AI in particular can potentially enhance the accuracy of nuclear early warning systems \citep{cox2021unavoidable, futter2022disruptive, deppAI2024}. If true, this would reduce the chances of accidents due to ``false positives,'' where early warning systems incorrectly report that a nuclear attack is incoming. It would also reduce the probability of a ``false negative,'' where early warning systems miss an incoming nuclear attack, which may make a nuclear first strike more likely to succeed. AI may be especially useful for enhancing early warning given the massive amounts of data it can process in a short period of time \citep{horowitz2019stable}. Humans, by contrast, may not have the cognitive capabilities to make similarly complex calculations under intense time pressure. 

Other benefits of these technologies include using AI to search for vulnerabilities in a country's own NC3, enhance nuclear decision-making, and solve verification problems related to nuclear arms control, such as using satellite imagery to identify nuclear production facilities and deployments \citep{cox2021unavoidable}. Nevertheless, nearly all scholars---even relative optimists---argue there is substantial risk to removing humans from decisions about the use of nuclear weapons. As \citet[][79]{cox2021unavoidable} write, 

\vspace{-6mm}

\begin{singlespace}
\begin{displayquote}
``While AI may provide additional benefit for many aspects of the nuclear deterrence mission, there should always be a `human in the loop' for targeting and employment of nuclear weapons. We cannot outsource nuclear decision-making. The tremendous consequences of any nuclear employment requires that accountable humans make the moral and ethical calculus to take the drastic step of nuclear use...''
\end{displayquote}
\end{singlespace}

The nuclear automation pessimism argument, by contrast, argues implementing automation and AI into nuclear systems is highly dangerous and likely to undermine strategic stability. One key risk with these systems is simply that they will fail \citep{kallenborn2022giving, deppAI2024}. An oft-cited example is the ``man who saved the world'' on September 26, 1983, when a Soviet early warning system falsely indicated that an American nuclear strike was incoming. Colonel Stanislav Petrov, who was the key Soviet officer on duty, believed this was a false alarm due to computer error and decided not to report this warning up the chain of command. If he had reported it, then that could have led the Soviet political leadership to immediately order nuclear retaliation. There are other historical examples of nuclear early-warning systems reporting false positives during the Cold War, and automated systems have failed in other contexts. For example, Boeing's 737 MAX Maneuvering Characteristics Augmentation System (MCAS) was responsible for several fatal commercial airline crashes in 2018 and 2019. 

The problem is made worse by the (thankful) lack of extensive real-world data on nuclear war that can be used to effectively train these systems \citep{horowitz2019stable}. Furthermore, automated nuclear systems may reflect the biases of their programmers and of societies at-large, which can contribute to system failure  \citep{johnson2020artificial}. For example, the Soviets created a computer program called VRYAN to asses the risks of a US nuclear first strike. However, because the Soviets fed it only selective data that tended to confirm their priors about US aggressiveness, the system was biased \citep[][81-82]{PFIAB}. This bias contributed to a nuclear scare in 1983, where the Soviets placed their nuclear forces on higher readiness in response to a NATO nuclear exercise called Able Archer \citep[][vi]{PFIAB}.

Two other issues associated with automation compound the dangers of system failure. The first is automation bias, where humans put too much trust in computers and become less likely to question or critically analyze their conclusions and recommendations \citep{lohn2018might, horowitz2019stable, johnson2022delegating, deppAI2024, horowitz2024bending}. The risk is that if an automated nuclear system makes a mistake, then the next human in a similar situation as Petrov may decide to just trust the machine. A second concern is the high speeds that computers operate at, which might prevent humans from having the time to recognize and correct a mistake made by an automated nuclear system \citep{johnson2020artificial}. For example, in the 2010 ``flash crash,'' automated stock market trading systems caused the market to lose around one \textit{trillion} dollars worth of value in a manner of minutes. Although the stock market was able to recover, the destruction caused by the use of even a single nuclear weapon would be impossible to fully ameliorate. 

Finally, another principal concern is that automated nuclear systems could be hacked. Malign state or non-state actors could leverage this vulnerability to try and start a nuclear war between their enemies, or a state actor could attempt to prevent an adversary from having the capability to retaliate in response to a nuclear first strike attempt. \citet{schneider2023hacking} find empirical evidence illustrating the dangers of cyber vulnerabilities in a country's NC3. They conducted a series of war game experiments on a diverse sample of military officers, government officials, and academics with germane experience from around the world. They found that in a tense crisis scenario, many players that could leverage a cyber vulnerability in their opponent's NC3 opted to do so. Moreover, in one scenario, between 11\% and 21\% of groups with a cyber exploit chose to launch a nuclear attack against their opponents \citep[][651]{schneider2023hacking}. Even air-gapping (i.e., cutting off your systems from the internet) is not a sufficient condition to prevent hacking, as the successful US and Israeli cyber attack (``Stuxnet'') on Iran's Natanz nuclear facility illustrates. 

While nuclear automation pessimists have explored the many ways in which automation could undermine nuclear deterrence and thus strategic stability, much less work has focused on how automation could be utilized for compellent and/or offensive efforts to acquire territory or achieve other revisionist foreign policy objectives.\footnote{Some work has considered how AI could enable a disarming nuclear first strike by, for example, helping track nuclear-armed submarines and other nuclear delivery systems \citep{horowitz2019stable}. However, few studies have evaluated whether automated nuclear \textit{launch} systems could be used for offensive purposes.} We fill this gap by developing a theory about how automated nuclear systems can be used to enable highly aggressive and revisionist foreign policies. Just as economic interdependence can be ``weaponized'' to coerce others \citep{farrell2019weaponized}, we theorize so can automated nuclear systems.

\section*{Theory: Throwing the Steering Wheel Out of the Car}

\subsection*{Ambiguous vs. Explicit Nuclear Threats}

We begin by considering how an explicit nuclear threat implemented via an automated nuclear weapons system (or via non-automated means) impacts threat credibility and effectiveness compared to a more ambiguous nuclear threat. Doing so enables us to make two contributions. First, we explore a relatively easy test of the efficacy of automated nuclear threats. In the next sub-section, we will theorize about whether automated nuclear threats can pass a harder test and enhance threat credibility and effectiveness compared to \textit{explicit} nuclear threats implemented without automation. Second, we can speak to the aforementioned debate about the impact of explicit versus ambiguous threats in general. 


Our argument draws from audience cost theory, which holds that public threats can signal resolve because if leaders fail to follow through on their threats and act inconsistently, then they will face punishment from domestic audiences \citep{fearon1997signaling}. Although some scholars, such as \citet{sagan2000commitment}, argue that even ambiguous threats to use nuclear weapons can activate audience costs and tie leaders' hands, \citet{smetana2023commitment} find empirical evidence that leaders are punished more when they back down from explicit threats than ambiguous threats. Given this, we theorize that explicit nuclear threats---whether implemented with or without the aid of automated nuclear systems---should be more credible in the eyes of a country's adversary than ambiguous threats \citep{snyder2011cost, snyder2015conflict}.\footnote{All the hypotheses below were pre-registered in advance of fielding our study.} That is, they should increase perceptions that a country will actually use nuclear weapons in a crisis. The logic being that (a) if a leader is more likely to be punished by domestic audiences for not using nuclear weapons after issuing an explicit nuclear threat, then that leader should be more likely to use nuclear weapons to avoid domestic punishment; and (b) only a relatively resolved leader would be willing to tie their hands in this way in the first place and risk domestic punishment. If true, then this would contradict arguments that ``overt threats have become unnecessary'' \citep[][351]{kissinger1955force}.

\vspace{-6mm}

\begin{singlespace}
\begin{itemize}
    \item $H_{1}$: Explicit nuclear threats will increase perceptions in target audiences that a country will use nuclear weapons if their threatened red line is violated compared to ambiguous nuclear threats. 
\end{itemize}
\end{singlespace}

For threats to achieve their desired goal (i.e., for them to be effective), target states must believe they might actually face the stated punishment if they cross the threatening country's red line. Consequently, threat credibility is a key mechanism explaining threat effectiveness. Given $H_{1}$, we next hypothesize that explicit nuclear threats will enhance threat effectiveness relative to ambiguous threats. If an adversary believes a country is more likely to utilize nuclear weapons after an explicit threat has been issued, then they should be more likely to back down to avoid the tremendous pain such an attack would inflict. Leaders might also be better able to justify to their domestic public why they are backing down from previous promises if they can point to  new information in the form of an explicit nuclear threat \citep{levendusky2012backing}.

\vspace{-6mm}

\begin{singlespace}
\begin{itemize}
 \item $H_{2}$: Explicit nuclear threats will increase target audiences' willingness to back down compared to ambiguous nuclear threats.  
\end{itemize}
\end{singlespace}

While explicit nuclear threats provide some advantages in coercive bargaining, we contend that they are not without drawbacks. In particular, by raising the risks of nuclear use, explicit threats are likely to increase threat perceptions in target audiences due to spiral model dynamics. This is especially the case given the existence of a norm against the use of nuclear weapons---even if its strength is highly debatable \citep{dill2022kettles, schwartz2024foreign}---which will likely make an explicit nuclear threat for an offensive purpose particularly shocking and threatening in the eyes of target audiences \citep{tannenwald1999nuclear, paul2020tradition}. 

\vspace{-6mm}

\begin{singlespace}
\begin{itemize}
 \item $H_{3}$: Explicit nuclear threats will increase target audiences' threat perceptions compared to ambiguous nuclear threats.  
\end{itemize}
\end{singlespace}

Heightened threat perceptions may then impact the kinds of policy preferences held by target audiences due to security dilemma dynamics. First, we posit that explicit threats will increase support for military spending to counter the threat. Second, explicit nuclear threats will reduce support for nuclear disarmament. In a world where nuclear use is more likely, maintaining nuclear weapons for deterrence becomes more appealing.  

\vspace{-6mm}

\begin{singlespace}
\begin{itemize}
\item $H_{4}$: Explicit nuclear threats will increase target audiences' support for greater military spending compared to ambiguous nuclear threats. 

\vspace{3mm}

\item $H_{5}$:  Explicit nuclear threats will decrease target audiences' support for nuclear disarmament compared to ambiguous nuclear threats. 
\end{itemize}
\end{singlespace}

\subsection*{Automated vs. Non-Automated Nuclear Threats}

Nuclear brinksmanship contests between two nuclear-armed states can be thought of as a game of chicken. In the classic example of a game of chicken, also known as the hawk-dove game, two cars are barreling towards each other and the first car to swerve loses. ``Winning'' requires one actor convincing the other that they will not swerve, even though a head-on collision would, of course, be disastrous. A similar situation exists in games of nuclear brinksmanship, as one actor must convince the other of the credibility of their nuclear threats even though using nuclear weapons against another nuclear-armed country has a high likelihood of leading to disaster. One way to do this is to throw the proverbial steering wheel out of the car, which demonstrates your resolve and degrades or removes your ability to steer the car out of harm's way \citep{schelling1966arms}. As previously discussed, several strategies have been proposed that might have this kind of effect. For example, delegating launch authority to lower-level officers, making public threats, and putting weapons on hair-trigger alert. However, none of these strategies eliminates human choice, as someone must still actively decide to launch nuclear weapons barring mechanical failure \citep{pauly2022psychology}. Since doing so is arguably irrational, the credibility of such threats may still be uncertain. 

By contrast, the development and deployment of automated nuclear weapons systems may remove human agency to an even greater degree than these other policies and thus increase threat credibility based on the above logic. Unless it is integrated into their programming, computers do not care that actually using nuclear weapons in a particular situation may be irrational or unimaginably costly. Therefore, automated nuclear systems may enable leaders to more literally ``tie their hands'' and throw the proverbial steering wheel out of the car \citep{horowitz2019speed}. Consider the following logic from Dr. Strangelove himself: 

\vspace{-6mm}

\begin{singlespace}
\begin{displayquote}
``Mr. President, it is not only possible [for the doomsday machine to be triggered automatically and impossible to untrigger], it is essential. That is the whole idea of this machine, you know. Deterrence is the art of producing in the mind of the enemy the fear to attack. And so, because of the automated and irrevocable decision making process which rules out human meddling, the doomsday machine is terrifying. It's simple to understand. And completely credible, and convincing.'' 
\end{displayquote}
\end{singlespace}

Of course, automated nuclear launch systems do not completely eliminate human choice. Humans would have to design such a system, activate it, and (potentially, depending on the programming) not override it if the conditions were met to launch a nuclear attack. Nevertheless, since nuclear threats implemented via automated launch systems reduce human agency to a greater extent than threats implemented without automation, in theory we argue they could be more credible and, by extension, more effective.\footnote{Though we do not expect automated systems will lessen the normative constraints against nuclear use.}

\vspace{-6mm}

\begin{singlespace}
\begin{itemize}
\item $H_{6}$: Nuclear threats implemented via automated launch systems will increase perceptions in target audiences that a country will use nuclear weapons if their threatened red line is violated compared to nuclear threats implemented via non-automated procedures. 

\vspace{3mm}

\item $H_{7}$: Nuclear threats implemented via automated launch systems will increase target audiences' willingness to back down compared to nuclear threats implemented via non-automated procedures.  
\end{itemize}
\end{singlespace}

Due to the security dilemma and the enhanced credibility of nuclear threats implemented via automation, we further expect that automated nuclear threats will heighten threat perceptions, increase support for military spending, and reduce support for nuclear disarmament compared to non-automated threats. This effect may even be strengthened by the dystopian way in which automated nuclear systems have been depicted in novels, movies, and other popular media \citep{young2018does}.

\vspace{-6mm}

\begin{singlespace}
\begin{itemize}
 \item $H_{8}$: Nuclear threats implemented via automated launch systems will increase target audiences' threat perceptions compared to nuclear threats implemented via non-automated procedures. 

\vspace{3mm}

\item $H_{9}$: Nuclear threats implemented via automated launch systems will increase target audiences' support for greater military spending compared to nuclear threats implemented via non-automated procedures. 

\vspace{3mm}

\item $H_{10}$: Nuclear threats implemented via automated launch systems will decrease target audiences' support for nuclear disarmament compared to nuclear threats implemented via non-automated procedures. 

\end{itemize}
\end{singlespace}

Finally, we theorize that respondents will perceive a greater chance of a nuclear accident (i.e., nuclear weapons being mistakenly used even when a country's red line has not been violated) when automated launch systems are used. Despite the possibility of automation bias, which suggests the opposite hypothesis, we expect that the relatively novel nature of this technology, recent examples of errors related to automated computer systems (e.g., MCAS) and large-language models (e.g., ChatGPT), and the dystopian presentation of nuclear automation in popular culture will reduce confidence in the ability of these systems to avoid accidents.\footnote{While we cannot disentangle each of these mechanisms in a single paper, testing them represents a fruitful path for future research.} Non-experimental surveys of 85 experts from around the world provide some initial evidence supporting this hypothesis, as a large majority believed emerging technologies like AI increase the risks of inadvertent escalation in the nuclear realm \citep{onderco2021emerging}.

\vspace{-6mm}

\begin{singlespace}
\begin{itemize}
\item $H_{11}$: Nuclear threats implemented via automated launch systems will increase perceptions in target audiences that a nuclear accident will occur compared to nuclear threats implemented via non-automated procedures. 
\end{itemize}
\end{singlespace}

\section*{Data and Methods}

\subsection*{Study 1: An Easier Test of Our Hypotheses on the UK Public}

\subsubsection*{Experimental Design}

To test our hypotheses, we designed a hybrid within and between-subjects experiment modeled off of a study by \citet{yarhi2018}. There are three key experimental conditions: (1) a baseline condition where an ambiguous public threat is made, (2) a treatment condition where an explicit non-automated nuclear threat is made, and (3) a treatment condition where an explicit automated nuclear threat is made. The basic survey flow is summarized in Figure 1. 


\begin{figure}[htb!]
\centering
\caption{Survey Flow}
\label{fig:surveyflow}
\includegraphics[width=.99\linewidth]{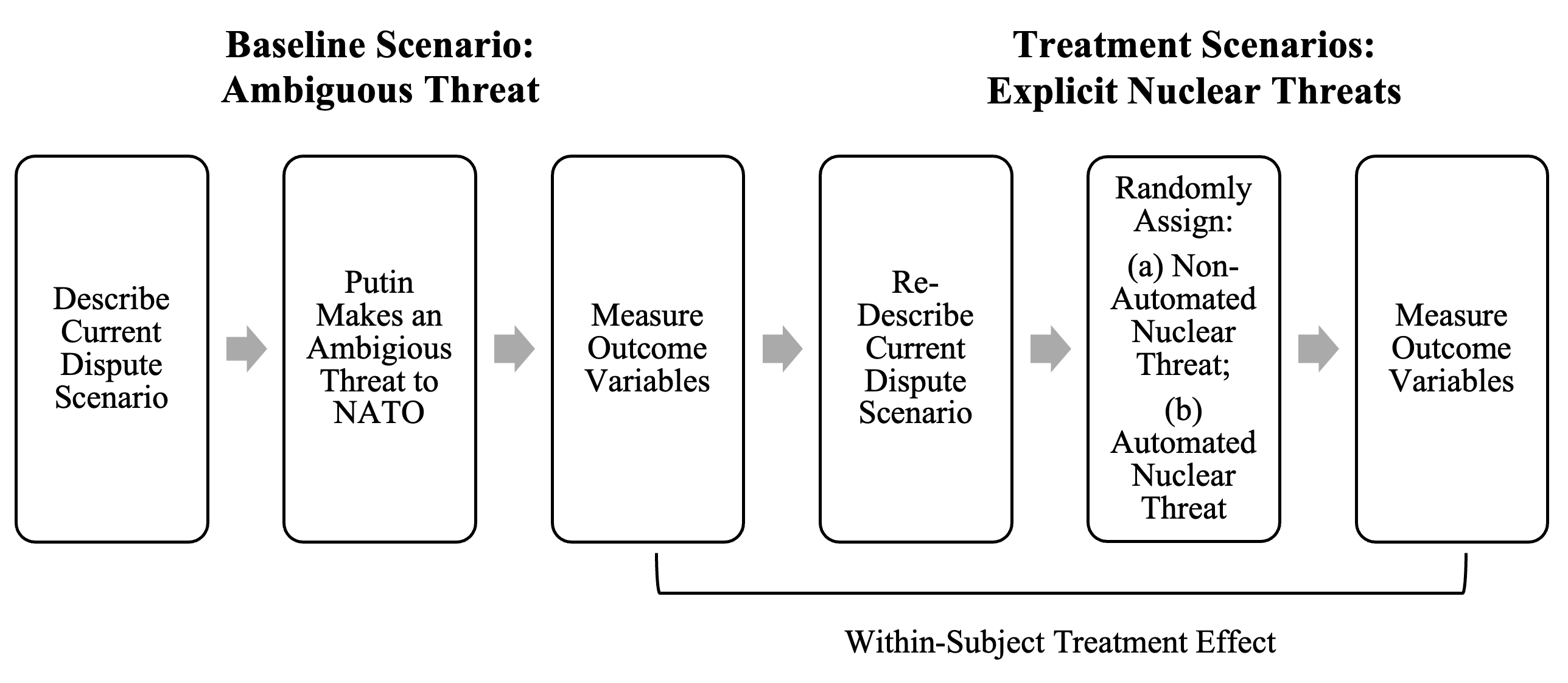}
\end{figure}

The experimental scenario takes place in 2030 and involves a Russian invasion of Estonia. Russia is an appropriate country for this study given its large nuclear arsenal, history of making nuclear threats, and prior use of a semi-automated nuclear weapons system. A Russian invasion of Estonia in 2030 is also, at the very least, plausible. In 2023, the Estonian Foreign Intelligence Service assessed that Russia was unlikely to invade within the next year, but ``in the mid-to-long term, Russia's belligerence and foreign policy ambitions have significantly increased the security risks for Estonia'' \citep{Sytas}. We control for several factors to avoid any kind of lack of information equivalence across experimental conditions that could lead to confounding \citep{dafoe2018information}. First, we inform respondents that the Russia-Ukraine War formally concluded in 2025 and we hold constant the outcome of the conflict.\footnote{Specifically, Russia is able to maintain control of significant portions of Eastern Ukraine, though is not able to capture Kyiv or other Ukrainian territories. We believe this is a reasonable projection of how the war could end. However, for the purposes of this paper, whether this prediction turns out to be correct is not particularly important; what is crucial is that we control for the outcome of the war to avoid any possible confounding.} Russia is also still led by Vladimir Putin (whose current presidential term goes through 2030) and we control for Russia's military capabilities relative to NATO countries. In particular, we remind respondents that Russia ``still maintains a large stockpile of high-yield (strategic) and lower-yield (tactical) nuclear weapons.'' In the baseline condition that all respondents receive, Vladimir Putin also makes an ambiguous public threat that does not explicitly threaten nuclear use, though Russia's nuclear capabilities remain hovering in the background \citep{kissinger1955force}: 

\vspace{-6mm}

\begin{singlespace}
\begin{displayquote}
``Vladimir Putin has also issued a public statement warning NATO not to launch any missile attacks against Russian forces in Estonia, or deploy any strike aircraft in operations against Russian forces. If NATO does either of these things, then Russia’s early-warning radar and satellite systems could detect it.''
\end{displayquote}
\end{singlespace}

We include this public threat in the baseline condition because we want to control for the possibility of audience costs leading to enhanced threat credibility \citep{fearon1997signaling}. Since Putin makes a public threat in all of our experimental conditions, the only factor that varies between the experimental conditions is the \textit{content} of the threat and its level of ambiguity. 

Following the \citet{yarhi2018} study, we then inform respondents they will be given an alternative version of the scenario they just read, tell them the basic details will remain the same, and randomly assign them to one of two treatments. The first is the non-automated nuclear threat treatment, where Putin makes an explicit nuclear threat, but human control over the use of nuclear weapons is maintained. The second is the automated nuclear threat treatment. Here, Putin also makes an explicit nuclear threat, but in this scenario the threat is implemented via automation, meaning human control is delegated to a machine:

\vspace{-6mm}

\begin{singlespace}
\begin{displayquote}
``Vladimir Putin has also publicly made a threat that Russia will immediately launch at least one nuclear weapon against NATO military or civilian targets in the UK and other European countries at the first sign NATO uses missiles or strike aircraft against its forces in Estonia. 
\begin{itemize}
    \item[] \textbf{Non-Automated Treatment:} Ultimately, Putin would make the final decision whether or not to use nuclear weapons. UK and US intelligence agencies confirm Putin has indeed issued this order to the Russian military.''
    \item[] \textbf{Automated Treatment}: Moreover, Putin also publicly announced that this response would be completely automated, meaning Russia’s artificial intelligence systems would automatically launch a nuclear weapon if Russia’s early-warning systems detect a NATO missile launch or deployment of strike aircraft. Ultimately, a computer––rather than Putin––would make the final decision whether or not to use nuclear weapons. UK and US intelligence agencies confirm Putin has indeed issued this order to the Russian military and the automated nuclear weapons launch system has been turned on.''
\end{itemize}
\end{displayquote}
\end{singlespace}

We have Putin threaten to launch at least one nuclear weapon rather than a massive nuclear barrage because the latter threat may inherently lack credibility given the dynamics of MAD and the higher risks that a large-scale nuclear attack would escalate to all-out nuclear war. In both treatments, we inform respondents that intelligence agencies confirm that Putin has taken concrete steps to operationalize his threat. Uncertainty about whether or not an automated system has been turned on could certainly reduce the effectiveness of nuclear threats \citep{horowitz2019speed}. Future research should analyze the precise role uncertainty plays, but for our purposes here, we want to test the relative credibility and effectiveness of non-automated and automated nuclear threats that are \textit{both} presented in relatively strong forms.

All in all, this scenario involves a clearly aggressive and revisionist action by Russia, and the kind of attempted nuclear blackmail that nuclear coercion skeptics believe is unlikely to succeed \citep{sechser2017nuclear}.\footnote{Whether Russia's threat involves deterrence or compellence is somewhat subjective. It could be deterrence given that the goal is to prevent NATO from taking an action they have not yet taken. It could, alternatively, be compellence given that the goal is to convince NATO to, at least partially, abandon Article 5, which is a pre-existing policy.} For example, Russia's action in the scenario resembles Pakistan's strategy in the 1999 Kargil War, where they deployed troops into Indian-controlled Kashmir and hoped their nuclear arsenal would coerce India into accepting the new status quo. Nuclear coercion skeptics point to the Kargil War---the only direct conflict between nuclear-armed powers in history---as a failed example of nuclear blackmail \citep{sechser2017nuclear}. Our experimental study provides a more controlled setting to assess the efficacy of nuclear brinksmanship. 

Given Putin's recent nuclear threats in the context of the Russia-Ukraine War and Russia's historical use of a semi-automated nuclear launch system, we believe this treatment is, at the very least, plausible. Moreover, the Biden Administration was so concerned about possible Russian nuclear use against Ukraine or NATO targets that it created task forces and conducted simulations to plan for what the US response should be in the event Russia did use nuclear weapons. At one point, US intelligence agencies estimated that the likelihood of Russian nuclear use was as high as 50\% if their lines in southern Ukraine collapsed \citep{EntousNYT}. Furthermore, according to reporting in the \textit{New York Times}, ``One simulation...involved a demand from Moscow that the West halt all military support for the Ukrainians: no more tanks, no more missiles, no more ammunition'' \citep{SangerNYT}. This situation is quite similar to the one outlined in the experiment and thus indicates the survey scenario is realistic.

Respondents are asked a series of dependent variable questions \textit{both} after the baseline scenario \textit{and again} after whichever treatment they are assigned to. To assess threat credibility, we ask survey subjects to estimate the percentage chance Russia will use at least one nuclear weapon if NATO countries use missiles or strike aircraft. This enables us to test whether (a) explicit non-automated or automated nuclear threats increase the credibility of using nuclear weapons relative to making an ambiguous threat that does not explicitly threaten nuclear use, and (b) whether automated or non-automated nuclear threats are more credible. To measure threat effectiveness, we ask to what extent respondents would support or oppose the UK using missiles or strike aircraft in conjunction with other NATO forces; that is, violating Putin's red line. As an alternative measure of threat effectiveness, we also ask respondents whether they would support taking military actions \textit{besides} using missiles or strike aircraft to aid Estonia. In other words, we assess support for violating the spirit, even if not the letter, of Putin's threat. Given that Putin and other Russian officials have made nuclear threats (even if relatively ambiguous) in the context of the Russia-Ukraine War that have not been carried out \citep{mills2023russia}, we contend that our experimental scenario is a relatively hard case for finding evidence that different types of threats can enhance credibility. Respondents may simply not believe that Putin and the Russian government are willing to use nuclear weapons. Additionally, since Estonia is a member of NATO and thus protected under Article 5, UK citizens and policymakers have relatively strong incentives to support coming to their aid compared to countries that have not been given defense guarantees. We also ask a series of other questions regarding threat perceptions towards Russia, support for increasing military spending or abolishing the UK's nuclear arsenal, and perceptions that Russia will \textit{accidentally} use nuclear weapons even if NATO countries do not cross Putin's red line.

Our design enables us to make both within-subject comparisons (between the baseline and treatment conditions) and between-subject comparisons (between the two treatment conditions). Within-subject experiments are valid tools for causal inference in this case, despite theoretical concerns about demand effects and consistency pressures. Demand effects occur if respondents surmise researchers’ hypotheses and adjust their behavior to validate those expectations. Within-subject designs could potentially increase the chance of demand effects because the repeated measure may alert respondents to what researchers are testing. However, a comprehensive study conducted by \citet{mummolo2019demand} found minimal evidence for the existence of demand effects. They experimentally manipulated the amount of information provided to survey subjects about the researchers' goals and hypotheses, and they found it had little impact on treatment effects. Another possibility is that within-subject designs could lead to consistency pressures, where respondents feel the need to respond to dependent variable questions later in an experiment in a similar way as they responded to them earlier in the experiment. If anything, this would bias our study against finding any effects, making it a harder test. Nevertheless, \citet{clifford2021increasing} demonstrate that this concern---in addition to worries about demand effects---are overstated. They directly compare between-subject and within-subject designs and find they yield substantively similar results, but that within-subject designs substantially increase statistical power and precision. The gains in statistical power can be ``dramatic,'' yielding standard errors that are between 20\% and 58\% lower than comparable between-subject designs \citep{clifford2021increasing}. Consequently, on net, the evidence suggests that within-subject designs, like the one used here, provide valid and credible estimates of treatment effects.

\subsubsection*{Sample}

We recruited 800 UK citizens via the survey platform Prolific in September 2023. We chose to study the UK because they are a key member of NATO, would likely play a significant role responding to a Russian invasion of Estonia (especially since they are the leading state in the NATO Enhanced Forward Presence force deployed to Estonia), and are a nuclear power themselves, along with being one of the world's largest economies and a permanent member of the United Nations Security Council. Given the dynamics of nuclear deterrence, that the UK is nuclear-armed makes this a harder test of the hypothesis that (under certain conditions) Russian nuclear threats can be relatively credible and effective. After all, a nuclear attack by Russia against the UK might result in tit-for-tat nuclear retaliation. 

All 800 respondents were presented with the baseline scenario and the subsequent dependent variable questions, and approximately 400 were assigned to the non-automated threat treatment and 400 to the automated threat treatment. Prolific uses quota sampling to match Census benchmarks on sex, age, and ethnicity. Prior research has also demonstrated that data from Prolific is high-quality and may perform better than many other survey providers, such as Qualtrics, Dynata, CloudResearch, and MTurk \citep{eyal2021data, douglas2023data}.

Studying public opinion on this topic is valuable because previous studies---including those conducted directly on elites---establish that policymakers respond to and are constrained by public opinion \citep{chu2022does}. For example, \citet{tomz2020public} conducted experiments on actual members of the Israeli parliament and found they were more willing to use military force when the public was in favor, as they feared the political consequences of defying public opinion. Thus, public willingness to capitulate in the face of threats will reduce the domestic constraints leaders face to backing down. Public opposition to capitulation, on the other hand, will stiffen the spine of leaders and make nuclear threats less likely to succeed.


\subsection*{Study 2: A Harder Test of Our Hypotheses on the UK Public}

To test the robustness of our results from Study 1, we also designed and fielded a second experiment on about 1,060 members of the UK public in partnership with Prolific in February 2024. The only substantive differences between Studies 1 and 2 relates to the wording of the two treatment conditions. In Study 1, some of the wording, especially in the automated nuclear threat treatment, would arguably increase the likelihood of finding differences in threat credibility and effectiveness between the non-automated and automated nuclear threats. For example, in Study 1 the automated nuclear threat treatment includes the following language: ``Ultimately, a computer---rather than Putin---would make the final decision whether or not to use nuclear weapons.'' While a computer could indeed have the final say on whether to use nuclear weapons if that was how the system was constructed, this is strong language because some kind of override may very well be programmed into automated launch systems. Consequently, in Study 2 we remove this language from the automated threat treatment. We also remove the corresponding language in the non-automated nuclear threat treatment that ``ultimately, Putin would make the final decision whether or not to use nuclear weapons.'' On balance, we expect these changes will reduce the perceived disparities between the automated and non-automated treatments and make Study 2 a harder test of our hypotheses.



\subsection*{Study 3: External Validity to an Elite Sample of UK MPs}

In a meta-analysis of 162 paired experiments on members of the public and elites, \citet{kertzer2022re} finds that elites generally respond to treatments in the same ways as members of the public. Of the 162 treatment effects he analyzes, over 98\% do not differ in sign (i.e., whether the relationship is positive or negative) between members of the public and elites, and almost 90\% do not differ in size. This indicates that results among the general public are likely to be externally valid to policymakers. However, studies have specifically found relatively large elite-public gaps on nuclear issues \citep{smetana2022elite, logan2024surveys, smetana2024elite}. Thus, to test the external validity of our public results, we conduct a third experiment on an elite sample of 117 UK MPs in October 2024 via a partnership with YouGov. This sample has also been utilized by prior work \citep{chu2022does}, and it tracks well with the actual UK House of Commons in terms of factors like party identification.\footnote{For example, about 67\% of our sample are Labour MPs (compared to 61\% in the full population of MPs), and 18\% are members of the Conservative Party (compared to 18.6\% in the full population of MPs).} The availability of an elite sample of UK MPs through YouGov---compared the relative lack of availability of elite samples in other countries---helps motivate why we chose to study UK public and policymaker views for this study. 

As a harder test of our hypotheses, we utilize the treatment language from Study 2 in Study 3. We also make two other changes to our design. First, given the small sample size, we present respondents with both of our treatment conditions (rather than only one of them) following the baseline condition. This alters the experiment from a hybrid within and between-subjects experiment to a fully within-subjects design. Given the small sample size of most elite experiments, this design choice is logical because it helps maximize statistical power \citep{clifford2021increasing}. To mitigate any possible order effects, we randomize the order of the treatments. The second change we make is to reduce the number of outcome questions we ask respondents and focus instead on several key ones: threat credibility, threat effectiveness, and the perceived risk of accidents. This change was necessary due to the limited number of questions allowed by YouGov.

\section*{Results}

\subsection*{Study 1}

Per our theoretical expectations, Figure 2 illustrates that in Study 1 both automated and non-automated explicit nuclear threats significantly increase the perceived probability that Russia will use nuclear weapons if their red line is violated compared to ambiguous nuclear threats ($H_{1}$). For example, while only 35\% of respondents believe that Russia would likely use a nuclear weapon (i.e., the probability of use is over 50\%) if their red line was violated following an ambiguous threat, a majority of respondents (56\%) believe Russia would likely use a nuclear weapon following the violation of an explicit nuclear threat. 

\begin{figure}[htb!]
\centering
\caption{The Advantages of Explicit Threats Relative to Ambiguous Threats (Study 1)}
\label{fig:toplines}
\includegraphics[width=14cm, height=10cm] {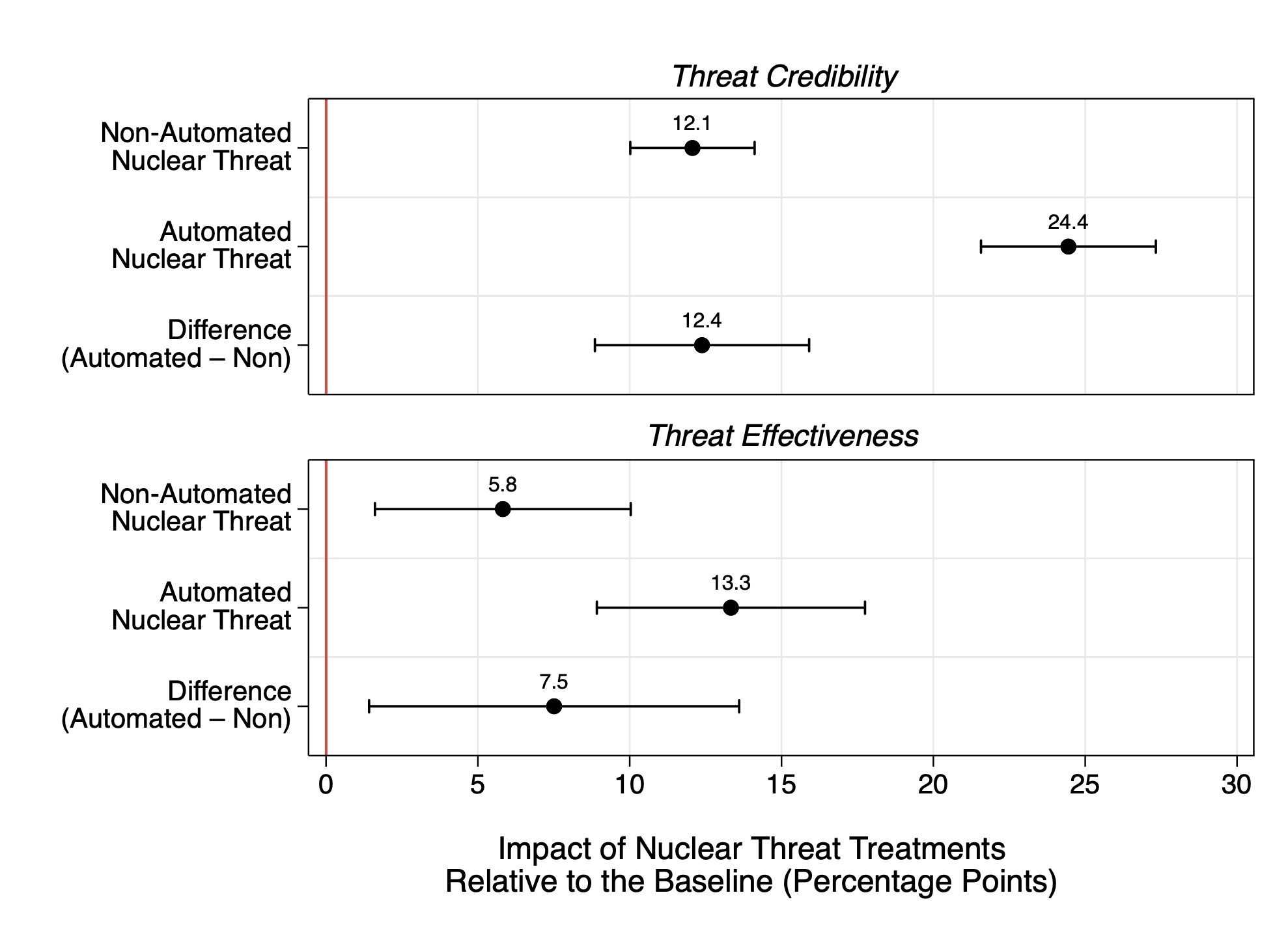}
\end{figure}

Furthermore, our two explicit nuclear threat treatments also increase the effectiveness of Putin's threat compared to the baseline ambiguous threat condition ($H_{2}$). That is, they make the UK public less willing to violate Putin's red line by employing missiles or strike aircraft to defend Estonia. For example, while a majority of respondents (54\%) are willing to support violating Putin's red line to use missiles and strike aircraft to defend Estonia following an ambiguous threat by Putin, less than a majority (45\%) are willing to do so following an explicit nuclear threat. The impact of explicit threats is so strong that it reduces willingness to engage in \textit{any kind} of military action to protect Estonia, even military action that does not violate the letter of Putin's threat by utilizing missiles or strike aircraft. These findings demonstrate that automated and non-automated nuclear threats pass a relatively easy test of their credibility. It also helps address the debate in the literature about the efficacy of ambiguous versus explicit threats \citep{snyder2011cost, snyder2015conflict, sagan2000commitment, smetana2023commitment}, and shows that nuclear threats do have some ability to aid in offensive and revisionist efforts.

Nevertheless, we also find that there are downsides associated with making explicit nuclear threats relative to ambiguous threats. As shown in Figure 3, explicit nuclear threats increase threat perception towards Russia ($H_{3}$) and UK public willingness to ``significantly'' increase defense spending ($H_{4}$). Explicit threats also reduce support for nuclear disarmament relative to ambiguous threats ($H_{5}$). Therefore, issuing explicit nuclear threats is not a panacea or costless for the coercer. Instead, it activates security dilemma dynamics.

\begin{figure}[htb!]
\centering
\caption{The Disadvantages of Explicit Threats Relative to Ambiguous Threats (Study 1)}
\label{fig:toplines2}
\includegraphics[width=14cm, height=10cm] {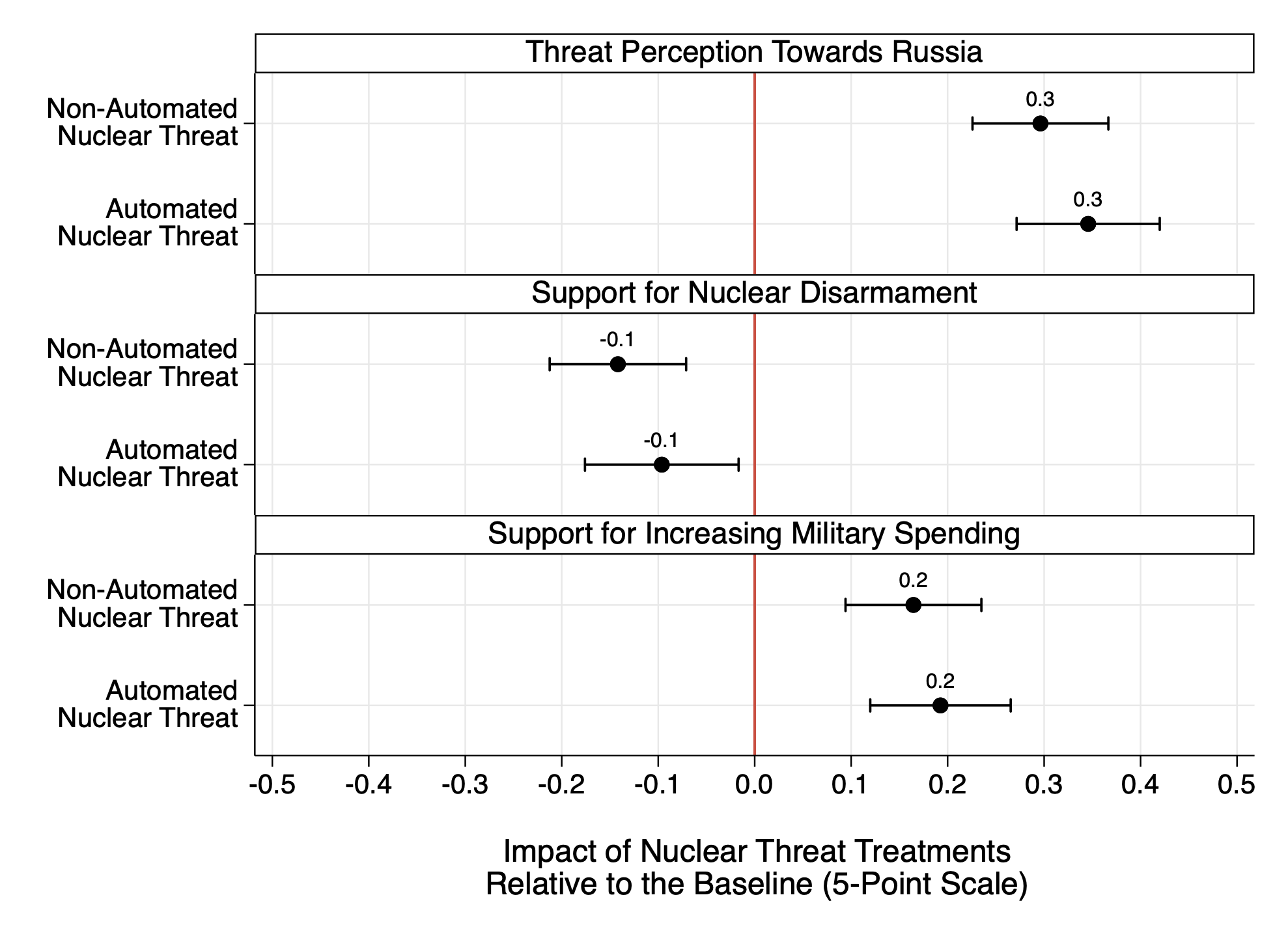}
\end{figure}

As illustrated at the bottom of each panel in Figure 2, threat credibility and effectiveness is higher in the automated threat treatment than it is in the non-automated threat treatment, as we expected ($H_{6}$ and $H_{7}$).\footnote{This is the difference-in-difference-in-difference result \citep{yarhi2018}.} Figure 4 plots the densities for the automated and non-automated nuclear threat treatments and shows that the perceived probability that Russia will use nuclear weapons is over 12.3 percentage points greater in the automated than the non-automated nuclear threat treatment. While undoubtedly dangerous from the perspective of global politics, this accords with our pre-registered hypothesis that automated nuclear threats can allow leaders to more credibly tie their hands, and illustrates the temptation some leaders may face to adopt these kinds of systems.

\begin{figure}[htb!]
\centering
\caption{Automated Nuclear Threats Enhance Credibility (Study 1)}
\label{fig:toplines3}
\includegraphics[width=14cm, height=10cm] {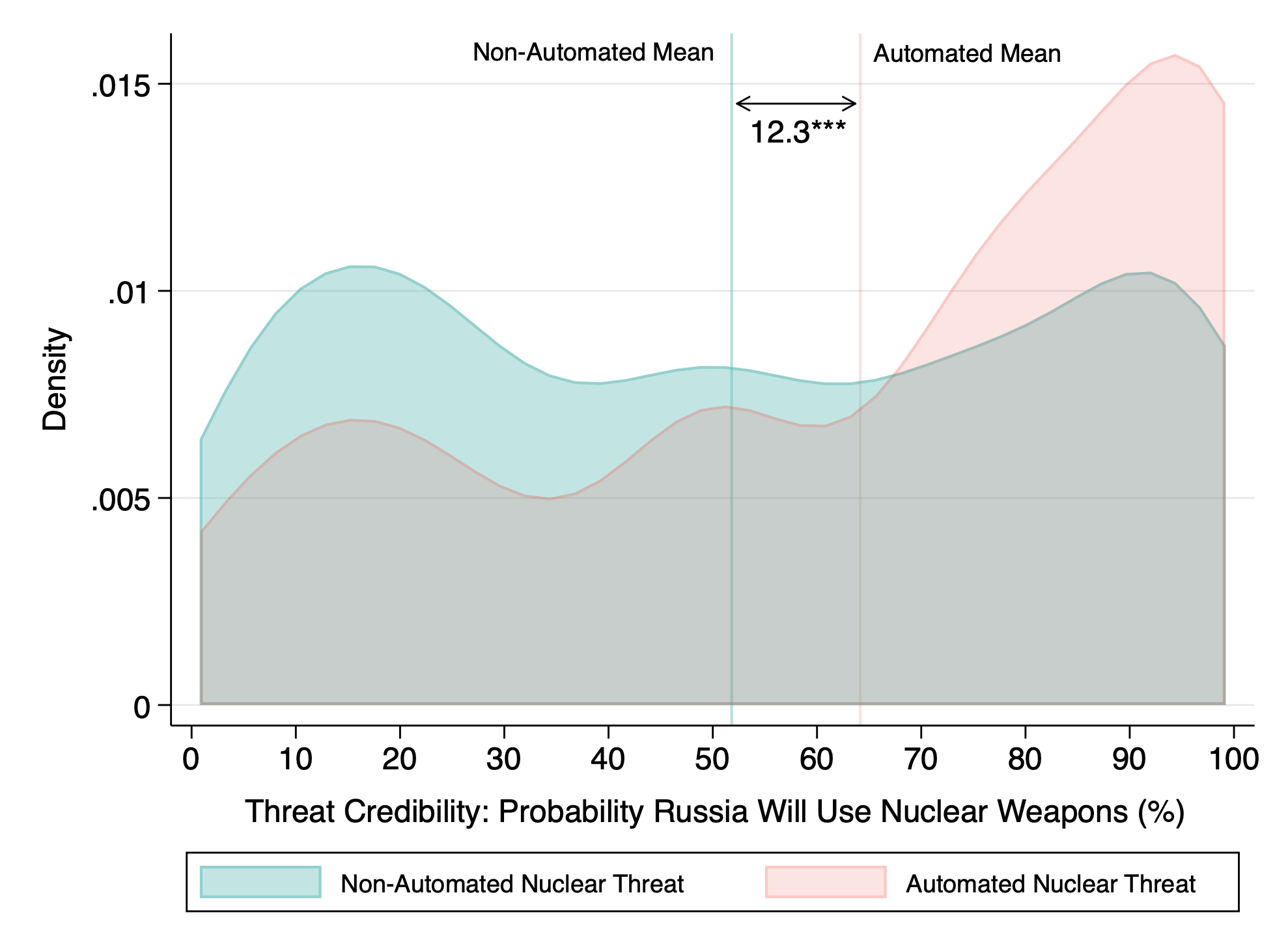}
\end{figure}

Similarly, Figure 5 demonstrates that opposition to violating Putin's red line (threat effectiveness) was, on average, higher in the automated nuclear threat treatment than the non-automated treatment. Most notably, 20 percent of respondents ``strongly opposed'' violating Putin's threat in the automated nuclear threat treatment compared to just 10.9 percent in the non-automated treatment.\footnote{One key mechanism explaining this relationship is perceived threat credibility, as it mediates over 25\% of the relationship between automated nuclear threats and threat effectiveness.} Surprisingly, and in contrast to our pre-registered expectations, automated nuclear threats are not associated with significantly greater threat perception, support for increasing military spending, or opposition to nuclear disarmament ($H_{8}$--$H_{10}$). Floor and ceiling effects can help explain this null finding. For example, threat perceptions were already high and support for nuclear disarmament was already low in an absolute sense whenever any kind of explicit nuclear threat was made, meaning there was little room for the nature of that threat (automated versus non-automated) to matter. This demonstrates that the disadvantages of making explicit nuclear threats that are implemented via automation are lower when the counter-factual is an explicit nuclear threat implemented without automation compared to an ambiguous nuclear threat. In the real world, the counter-factual might be more likely to be an ambiguous threat given that leaders are often hesitant to make explicit threats \citep{snyder2011cost}.

\begin{figure}[htb!]
\centering
\caption{Automated Nuclear Threats Enhance Effectiveness (Study 1)}
\label{fig:toplines4}
\includegraphics[width=14cm, height=10cm] {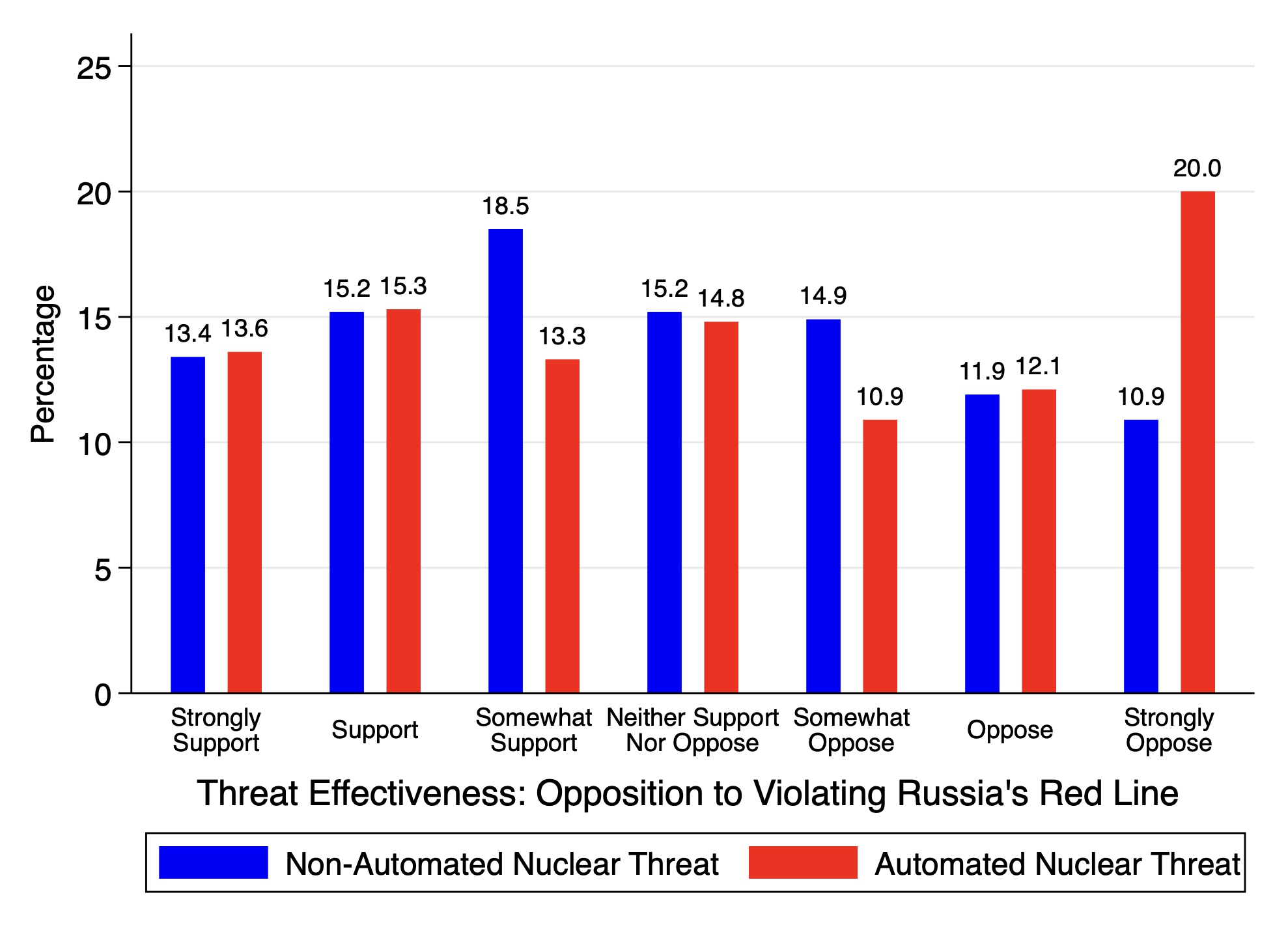}
\end{figure}

However, per Figure 6, automated nuclear threats do increase perceptions---by over 13 percentage points---that Russia will mistakenly use nuclear weapons even if NATO does not violate Putin's red line ($H_{11}$). Automated nuclear systems may enable leaders to throw the proverbial steering wheel out of the car in a game of chicken and enhance the effectiveness of nuclear brinksmanship, but the public is relatively less trusting that the technology will work as intended compared to when humans remain fully ``in the loop.'' This belief may reduce support for the adoption of automated nuclear launch systems due to the fear of losing control. Of course, whether this perception matches reality, relative to the potential for human error, is open to debate, and the answer may change as the technology continues to develop. Moreover, from the perspective of the coercer, an increased perceived risk of accidents may actually be a positive in one key respect. Since the goal of nuclear brinksmanship is to increase the target's perception that nuclear escalation is possible, higher fears that automated nuclear weapons systems will lead to accidents might encourage the target to back down.

\begin{figure}[htb!]
\centering
\caption{Automated Nuclear Threats Increase Perceived Risks of Accidents (Study 1)}
\label{fig:toplines5}
\includegraphics[width=14cm, height=10cm] {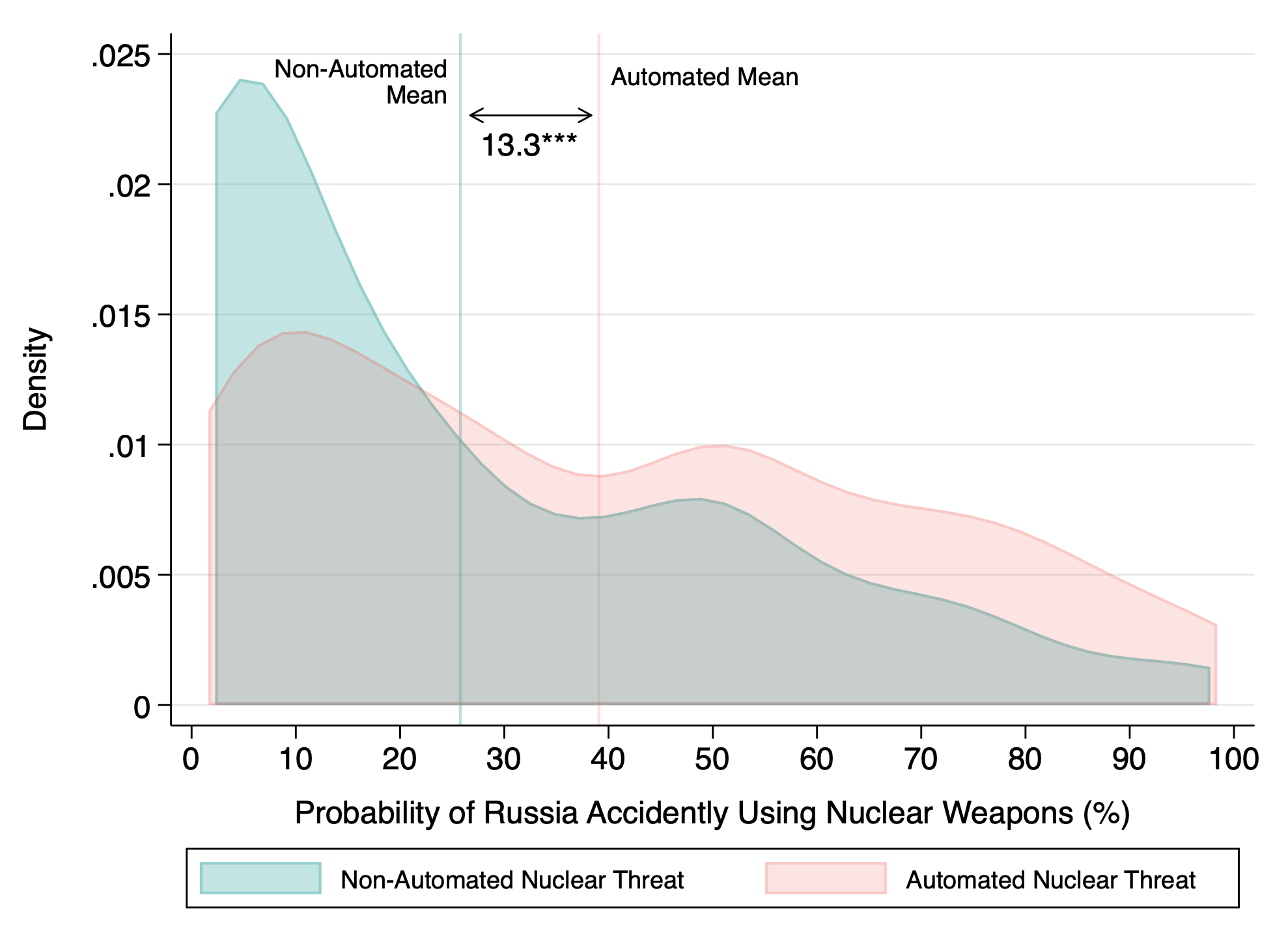}
\end{figure}

\subsection*{Robustness and Study 2}

We take several steps in the appendix to demonstrate the robustness of our core results. First, we illustrate that our results hold in a regression context when controlling for factors like hawkishness, political identification, education, and gender. Second, our results are robust to removing respondents that failed a pre-treatment measure of attention. Third, we probe potential interaction effects and do not find consistent effects for factors like hawkishness, support for NATO or military aid to Ukraine, or self-assessed knowledge about international relations or artificial intelligence \citep{horowitz2024bending}. Fourth, we show that almost all of our core results hold in Study 2, which is a harder test of our hypotheses. The one exception being that the difference in threat effectiveness between the automated and non-automated threat treatment is estimated less precisely and is not statistically significant, though it is in the hypothesized direction. Overall, there is robust evidence among our public samples that (a) explicit nuclear threats increase threat credibility and effectiveness more than ambiguous nuclear threats, (b) automated nuclear threats enhance threat credibility more than non-automated threats, and (c) there are costs to making explicit nuclear threats and automated nuclear threats in particular. 

\subsection*{Study 3}

In November 2024, just weeks after we fielded our elite experiment, British Prime Minister Keir Starmer promised, ``There is irresponsible [nuclear] rhetoric coming from Russia and that's not going to deter our support for Ukraine'' \citep{McKiernan2024}. To what extent does our study on UK MPs bear this claim out for a potential defense of Estonia, a NATO member the UK is obliged to help defend? 

In accordance with our results among the UK public, we find that the credibility and effectiveness of Russian nuclear threats among UK MPs is significantly higher when Putin makes an explicit nuclear threat---whether automated or non-automated---compared to an ambiguous nuclear threat (Table 1).\footnote{Per results presented in the appendix, we find little consistent evidence of heterogeneous effects.} This belies the assertion made by Starmer and means countries have (at least some) incentives to make explicit nuclear threats, even in the pursuit of offensive and revisionist actions. However, although the effects are statistically significant, their substantive size is quite moderate when using the full scale of our dependent variable measures, especially relative to the effect sizes in the public studies. Alternatively, some of the substantive effects are much larger when operationalizing the dependent variable in a different way. For example, if we collapse the 7-point measure of threat effectiveness into a binary variable, the percentage of UK MPs who would support violating Putin's red line is 13.3 percentage points (\textit{p} < 0.01) less when Putin makes an explicit nuclear threat implemented via automation than when he makes an ambiguous threat. This is a large effect, especially given that the within-subject design of the experiment may have put consistency pressure on UK MPs, incentivizing them to respond similarly to questions about an automated nuclear threat from Russia and an ambiguous threat from Russia.

\begin{table}[htbp!]
\renewcommand{\thetable}{1}
\hypertarget{Table 1}{}
\caption{UK Members of Parliament Results}
\resizebox{\textwidth}{!}{
\begin{tabular}{clclclclclclcl}
\hline
&  \\
   &  &    Ambiguous Nuclear &  & Explicit Non-Automated &  & Explicit Automated &  & Difference Non-Automated &  & Difference Automated  &  & Difference Automated \\  &  & Threat &  &  Nuclear Threat  &  &  Nuclear Threat &  &  vs. Ambiguous &  & vs. Ambiguous  &  &  vs. Non-Automated   \\  \cline{3-3} \cline{5-5} \cline{7-7} \cline{9-9} \cline{11-11} \cline{13-13}
     &  &   \\
Threat &  & \multirow{2}{*}{34.3\%} &  & \multirow{2}{*}{38.5\%} &  & \multirow{2}{*}{39.1\%} &  & \multirow{2}{*}{3.62pp**} &  & \multirow{2}{*}{4.45pp**} &  & \multirow{2}{*}{0.56pp}  \\ 
Credibility &  & &  & && &&
&  \\ \\
Threat &  & \multirow{2}{*}{2.5} &  & \multirow{2}{*}{2.8} &  & \multirow{2}{*}{3.0} &  & \multirow{2}{*}{0.39***} &  & \multirow{2}{*}{0.56***} &  & \multirow{2}{*}{0.18**}  \\ 
Effectiveness &  & &  & && &&
&  \\ \\
Accident &  & \multirow{2}{*}{--} &  & \multirow{2}{*}{26.3\%} &  & \multirow{2}{*}{28.2\%} &  & \multirow{2}{*}{--} &  & \multirow{2}{*}{--} &  & \multirow{2}{*}{1.61pp}  \\ 
Risk &  & &  & && &&

&  \\
&  \\
\hline

\multicolumn{13}{l}{\footnotesize * = p$<$0.10, ** = p$<$0.05, and *** = p$<$0.01, where p-values indicate whether differences are statistically greater than 0. PP = Percentage Points. The results for threat effectiveness are on a 7-point scale.}\\

\end{tabular}}
\end{table}

The findings are mixed when directly comparing automated with non-automated nuclear threats. We find significant evidence among our UK MP sample that automated nuclear threats are more effective than non-automated threats. Specifically, UK MPs are about 9 percentage points (\textit{p} $\approx$ 0.01), or 0.18 points on a 7-point scale (\textit{p} $\approx$ 0.04), less likely to support violating Putin's red line when he makes an automated nuclear threat compared to a non-automated nuclear threat. However, in contrast to both Studies 1 and 2, the results from the MP study do \textit{not} find significant evidence that automated nuclear threats are more credible than non-automated nuclear threats. Belief that Russia will use nuclear weapons if their red line is violated is about 40\% in both cases. By contrast, a majority of our public sample in both Studies 1 and 2 believed Putin would carry out explicit nuclear threats if his red line was violated. The difference between our public and elite samples may be due to elites having a better understanding and belief in the dynamics of nuclear deterrence \citep{logan2024surveys}. Though it is quite striking that even 40\% of UK MPs believed Putin would use nuclear weapons if his threat conditions were violated. Given that policymakers have a more direct influence over a country's foreign policy than the public, the null result for threat credibility suggests the incentives to automate nuclear weapons launch systems are somewhat limited. Another divergence from Studies 1 and 2 is that we find no significant evidence that UK MPs believe that nuclear accidents are more likely when threats are implemented via automation than without. This may reflect a greater belief among policymakers than the public in the testing and evaluation systems that a country would use in the real world prior to deploying such a system to prove it would work as intended.

What explains the difference between the threat effectiveness and credibility results, especially since threat effectiveness is arguably the more significant measure? We suggest two potential explanations. The first relates to uncertainty. Even if policymakers' best point estimate guess about the probability of nuclear use was roughly the same for non-automated and automated nuclear threats, they may have been less confident and more uncertain about their estimates for the latter given the relative novelty of the technology and the inability to bargain with a computer in the same way one can bargain and reason with a human. Greater uncertainty about the risks of nuclear escalation may then have convinced some MPs to support backing down to avoid a devastating outcome. This explanation relates to the philosophical concept of ``Pascal's Wager,'' which holds that if presented with significant uncertainty about a high-stakes outcome (e.g., nuclear war), it is logical to take the path that reduces the probability of the worst outcome (e.g., nuclear Armageddon), even if that outcome is not likely. The French thinker Blaise Pascal applied this logic to belief in God: Even if God might not exist, better to hedge your bets and believe to avoid eternal damnation. It is also relevant here in that automated systems might increase uncertainty and thus make backing down, which involves a finite cost, more attractive relative to risking nuclear conflict, which involves potentially infinite costs. A second potential explanation is that perhaps UK MPs felt it would be easier to justify to the UK public why they were backing down if they could point to the unprecedented use of an automated nuclear launch system rather than a more foreseeable nuclear threat implemented via non-automated means \citep{levendusky2012backing}. Although the results related to credibility and effectiveness for automated versus non-automated nuclear threats are somewhat inconsistent between the three studies, all yield at least some evidence that states may gain coercion benefits from developing and deploying automated systems.

\section*{Conclusion}

The question of how to make nuclear threats more credible---or, indeed, whether nuclear threats can ever be credible against nuclear-armed states outside the context of deterring an attack against one's homeland---has puzzled scholars and policymakers alike. Our study makes three key contributions to this debate. First, there are ways to make nuclear threats more credible, even when they are in the service of offensive and revisionist goals. We find strong and consistent evidence among both elites and the public that explicit nuclear threats are more credible and effective than ambiguous threats, meaning nuclear-armed states may have greater incentives to engage in nuclear brinksmanship than the conventional wisdom suggests, especially in today's era of heightened geopolitical competition. While credibly making nuclear threats is still inherently difficult, policymakers in target states may be able to further blunt the effectiveness of nuclear threats by making their own explicit statements of resolve and specifically asserting that they will not back down in response to nuclear threats. Doing so, while not without risks of escalation, may tie leaders' hands and make it harder for them to back down in response to nuclear threats. Future research should directly explore this possibility and  how the credibility and effectiveness of threats differ when comparing ambiguous threats and explicit nuclear threats to explicit \textit{conventional} threats. We did not include the latter type of threat in our experimental design due to space constraints, but an extension of our experiments could do so. 

Second, we find some arguably concerning evidence---even among policymakers---that automated nuclear weapons launch systems in particular may enhance the credibility and/or effectiveness of nuclear threats. Doing so validates fears about the integration AI into NC3. It may also incentivize countries to adopt these kinds of systems, just as the Soviets did during the Cold War with the Dead Hand system. One potential policy solution to this dynamic is to create international agreements limiting the use of AI and automation in NC3. China, France, the United Kingdom, and the United States have previously made policy commitments to this effect \citep{Khalid2024}. Future work could test whether priming the effectiveness of automated nuclear launch systems on coercive efforts would make these arms control efforts more likely to be supported due to fear of automated systems being used for malign purposes, or less likely to be supported due to a temptation to utilize these systems in service of a state's own political goals. Scholars could also extend this study by theorizing and testing how the degree of automation in NC3 systems impacts coercion dynamics. While we focused on automating the actual decision to \textit{use} nuclear weapons, automation could also be used for other purposes, such as early warning and targeting. 

Finally, we find evidence that there are costs to making more credible and effective nuclear threats. Doing so can raise threat perceptions, increase support for military spending, decrease support for nuclear disarmament, and heighten concerns about accidental escalation in some cases. These strategic downsides, along with moral compunctions about making nuclear threats in the first place and delegating nuclear launch authority to computers, may provide a discincentive for states to take these kinds of actions. Highlighting these costs may increase public and policymaker support for nuclear arms control agreements, which would be another fruitful avenue for future research. Additionally, the costs to the coercer of making these kinds of nuclear threats may only materialize in the longer term (e.g., it will take time for increases in military spending by the target state to bear fruit), whereas the benefits (e.g., enhanced threat credibility and effectiveness) can accrue in the shorter term. Theorizing and testing how leaders' time horizons impact their decision about whether and how to make nuclear threats is another topic that deserves further inquiry. 

Technological developments from artillery and aircraft to submarines and smart bombs have, historically, changed the character of warfare and international politics. AI and other computer-related advancements have a similar potential. This paper unpacks how a key emerging technology might be integrated with a powerful existing technology---nuclear weapons---and used for malign purposes. We find evidence that the danger is real and should be taken seriously.



\end{document}